\documentclass[twocolumn,aps,showpacs]{revtex4}

\input epsf
\usepackage{graphicx}
\usepackage{bm}
\usepackage{multirow} 
\usepackage{verbatim} 
\usepackage{color}

\begin{document}

\title{Electronic band structure of 
platinum low--index surfaces: \\
an {\it ab initio \ vs.} tight--binding study. II}

\author{H. J. Herrera--Su\'arez}
\affiliation{Universidad de Ibagu\'e,
Carrera 22 Calle 67 Barrio Ambals\'a,
Facultad de Ciencias Naturales y Matem\'aticas,
Colombia
}
\author{A. Rubio--Ponce}
\affiliation{
Departamento de Ciencias B\'asicas, Universidad Aut\'onoma 
Metropolitana--Azcapotzalco, 
Av. San Pablo 180, 02200 M\'exico, D. F.}

\author{D. Olgu\'\i n}
\affiliation{ Departamento de F\'\i sica, Centro 
de Investigaci\'on y de Estudios Avanzados del 
Instituto Polit\'ecnico Nacional, 
A.P. 14740,
M\'exico, D.F. 07300 }



\begin{abstract}
As a second part of a previous paper, here the calculated
electronic band structure of ideal Pt(100) and Pt(110) 
surfaces, studied using density functional 
theory and the empirical tight--binding method, is presented.
A detailed discussion of the surface-- and resonance--states is given. 
It is shown that the calculated surface-- and resonance--states of 
ideal Pt(100) surfaces agree very well with the 
available experimental data.
For Pt(110), some of the surface-- and resonance--states are 
characteristic of the low degree of symmetry of the surface and 
are identified as being independent of surface reconstruction effects. 
As in the previous paper, 
the density functional calculations were performed using the full 
potential linearized augmented plane wave method, and the empirical 
calculations were performed using the tight--binding method and 
Surface Green's Function Matching Method.
\end{abstract}
\maketitle

\section{INTRODUCTION}

This work presents the continuation of our study 
of the Pt low--index surfaces.
In a previous paper we have discussed our calculations 
on Pt(111) surface \cite{JHerreraPt111}, and in this paper 
we will discuss our 
results on Pt(100) and Pt(110) surfaces.

The Pt(100) surface is usually studied in the 
$(1 \times 1)$ and $(5 \times 1)$ phases. 
The unreconstructed $(1 \times 1)$ phase is metastable, 
whereas the reconstructed $(5 \times 1)$ phase is obtained 
after the sample is annealed at 400 K \cite{APStampfl,RDrube,Subaran}. 
The reported surface-- and 
resonance--states of the metastable 
$(1 \times 1)$ phase \cite{APStampfl} 
were accurately reproduced in the present work.
The experimental reports of controversial surface--states
are clarified in this work.
A surface--state 
that was recently reported by Subaran {\it et al.} \cite{Subaran},
and was not observed 
in previous reports is properly identified in these calculations.
  
It is established that the Pt(110) surface exhibits the so--called 
$(1 \times 2)$ missing row reconstruction, whereas the 
$(1 \times 1)$ phase is metastable \cite{NMemmel,AMenzel,MMinca,GRangelov}. 
In this work, however, the calculation was performed on an ideal 
Pt(110) surface.
Although we did not find experimental data related to this phase, 
for completeness, we will discuss our results and will 
qualitatively compare our results with previous experimental 
data on the $(1 \times 2)$ missing row phase \cite{NMemmel, AMenzel}. 
These calculations reveal several surface--
and resonance--states 
that are reported to be characteristic of a low degree of symmetry 
of the surface. 
These states are identified as being independent of surface 
reconstruction effects, and
these facts support our approach to computations of the 
characteristics of this surface.
  
The rest of the paper is organized as follows: 
In Section \ref{metodo}, the 
essentials of the calculation methods are given.
Sections \ref{resultados}--\ref{Pt110} contain the results and a 
discussion of the studied surfaces.
A comparison of these results with ETB calculations is also 
presented in these sections.
Section \ref{conclusion} summarizes our work.
  
\section{COMPUTATIONAL Strategy}
\label{metodo}
  

The DFT calculations were done using the full potential 
linearized augmented plane wave (LAPW) method \cite{Blaha},
while the empirical tight--binding (ETB) calculations 
were done using the parametrization of 
Papaconstantopoulos \cite{Papa} and the 
surface Green function matching (SGFM) method 
of Garc\'ia--Moliner and Velasco \cite{Garcia-Moliner}. 
The details of the 
DFT and ETB
methods used in our calculations 
can be seen in Reference \cite{JHerreraPt111}.
Here we only comment the essentials of the 
DFT variational parameters used in our 
supercell calculations.
The ETB parameters and model used 
in the empirical calculations are the 
same as those discussed in Ref.~\cite{JHerreraPt111}.

To minimize the total energy
of the Pt(100) surface, a
supercell with 15--atomic layers and 9--vacuum layers was used, 
whereas a
supercell of 21--atomic layers and 13--vacuum layers was used 
for the Pt(110) surface
  
In the calculations, the step analysis was 
carefully performed to ensure the convergence 
of the total energy in terms of the variational parameters.
At the same time, an appropriate set of $k$ points 
was used.
The variational parameters used for the two studied surfaces
were $R_{\rm kmax}=9$ Ry and $G_{\rm max}= 14$.
The total energy of the Pt(100) surface was minimized using 
a set of 91 $k-$points in the irreducible portion of the BZ, equivalent
to a $(25 \times 25 \times 1)$ Monkhorst--Pack \cite{Monkhorst} grid
in the unit cell.
For the Pt(110) surface, the total energy was minimized using
a set of 88 $k-$points in the irreducible portion of the BZ, 
equivalent to a $(22 \times 16 \times 1)$
Monkhorst--Pack \cite{Monkhorst} grid. 
Finally, the total energy converged with a resolution better 
than 0.0001 Ry.
  
\begin{figure*}[!ht]
\vspace{3cm}
\includegraphics[width=120mm]{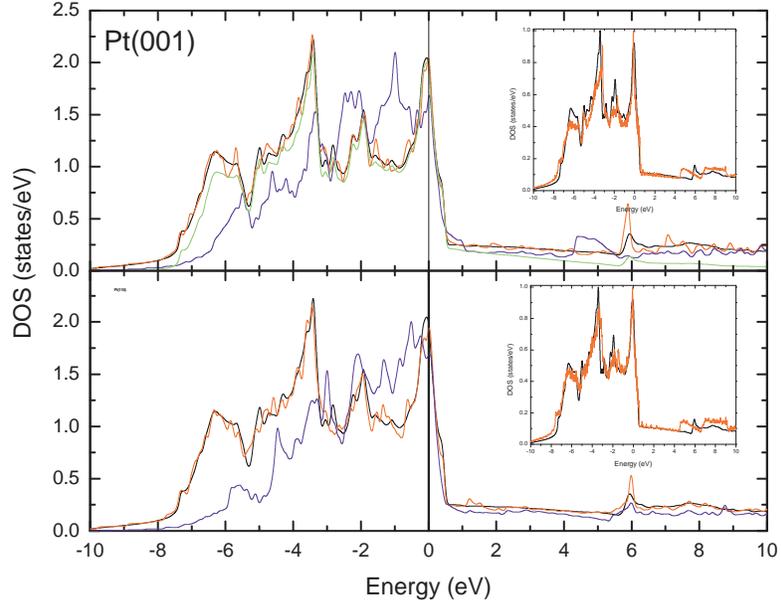}
\vspace{-3cm}
\caption{(Color online) Calculated DOS of the different
  Pt--surfaces studied in this work.
  The bulk DOS is presented as a black line, the DOS projected
  onto the central atomic layer is presented as a red line,
  and the DOS projected on the surface atomic layer is presented as
  a blue line.
  For comparison, the partial Pt--$5d$ contribution to the DOS 
  is also presented as a green line (upper panel).
  The inset in each panel shows the comparison of the 
  bulk DOS calculated using the FLAPW (black line) method and the 
  bulk projected DOS calculated using the SGFM--ETB (red line)
  method.
  }
\label{pt-dos}
\end{figure*}

As in the case of Pt(111) \cite{JHerreraPt111}, 
and to check the accuracy of the 
electronic properties calculated from the supercell approach,
the calculated bulk density of states (DOS) is compared with
the DOS projected 
onto the central atomic layers of the different 
supercells.
It should be noted 
that in this approach, the DOS projected onto the central atomic layer must be 
similar to the calculated bulk DOS.
Figure \ref{pt-dos} shows that this is the case.
In the figure, the calculated bulk DOS 
is shown as a solid line, the calculated DOS
projected onto the central atomic layer is presented as a dotted line,
and the calculated DOS projected onto the outer atomic
layer is presented as a broken line. 
In the upper panel, 
the partial bulk--DOS due to the Pt--$5d$ orbitals
is also shown.
The figure shows that 
in the energy range from approximately --7.0 eV to 0.5 eV, 
the main contribution to the bulk DOS is obtained from the $d$
electrons.
This symmetry composition should be 
reflected in the obtained surface electronic band structure.
The upper panel shows the calculated DOS of Pt(100)
and the lower panel shows the results for Pt(110).
In the figures, the zero of the energy axis represents the Fermi level $(E_F)$.
The figure shows that below $E_F$, the DOS projected onto the 
central layer (broken line) properly reproduces the main features of 
the bulk DOS (solid line) for each studied surface.
The bulk DOS exhibits four main peaks that are accurately reproduced 
by the DOS projected onto the central atomic layer. 
The same is true of the width and energy of the main peak.
The small observed differences are related to the shape of the main peaks.
Above $E_F$, Fig. \ref{pt-dos} shows that the DOS projected 
onto the central layer properly reproduces the bulk DOS up
to 6.0 eV, at which point some differences between 
the two calculations were observed.
  
However, it is clear from Fig.~\ref{pt-dos} that the
calculated DOS projected onto the outer atomic layer
is significantly different from the bulk DOS.
There are important features obtained from the projected surface DOS;
these features were obtained below and above $E_F$ but were not 
obtained for the bulk DOS.
Information about the 
surface-- and resonance--states will be found from these differences.
Below $E_F$, 
resonance--states are expected to be obtained, primarily because 
these energies represent 
the continuum of the projected bulk bands,
and few energy gaps exist at these energy values.
The surface--states will be obtained
above $E_F$ because energy gaps are more frequently
observed at these energies.

The comparison of the DOS calculated using the ETB and the 
DFT methods is shown in the inset of each figure.
As can be observed the calculated DOS using both methods are
quite similar, mainy for energies below $E_F$.
From these facts, it will be shown thta the found surface 
electronic band structure is also quite similr in both
calculations.

\section{Results and Discussion}
\label{resultados}

\subsection{Platinum(100)}
\label{Pt100}
  
\begin{figure*}[!ht]
\vspace{.75in}
\includegraphics[width=120mm]{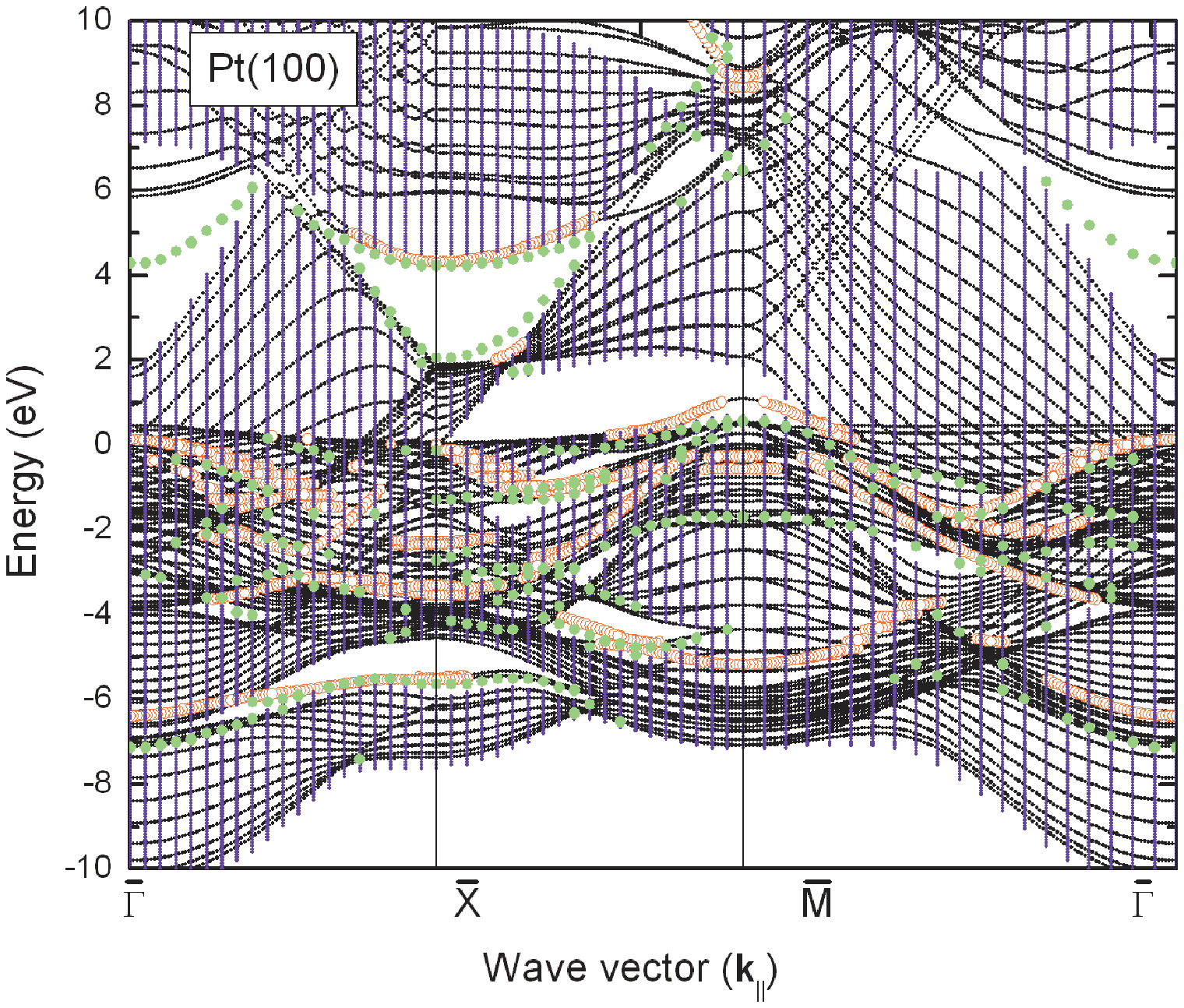}
\vspace{-3cm}
\caption{(Color online) Projected bulk bands of Pt(100).
  Black dots represent the DFT calculated pbbs.
  Red dots represent the SSs or the RSs if the states
  are located in a local energy gap or in the 
  continuum bulk bands, respectively.
  The blue dots represent the pbbs calculated by the 
  ETB method, and the SSs and RSs are represented 
  by green dots. 
}
\label{pt100}
\label{pt100-tb}
\end{figure*}

Figure \ref{pt100} 
shows the calculated DFT pbbs as well as the SSs and RSs of Pt(100).
Table \ref{tabla2} shows the wavefunction compositions
of the different SSs and RSs.

For this surface, at the $\bar X$ point,
an SS was found approximately 4.3 eV above $E_F$.
An RS was also found at the $\bar M$ point at energies that
range from 9.0 eV to 10.0 eV,
as seen in Fig. \ref{pt100}. 
These states are supported by the calculated DOS projected
onto the surface
as noted in Fig. \ref{pt-dos}.
According to the DFT calculations, the wavefunctions of these 
states have the symmetries of the $s,d_{x^2+y^2}$ and 
$d_{xy}$ orbitals, respectively.
  
However, 
as can be observed in Fig. \ref{pt100},
a number of SSs and RSs were obtained at energies below $E_F$.

According to the convention for a resonance state given above,
an RS was obtained at lower energies, approximately 6.4 eV below $E_F$.
The state seems to begin at the $\bar M-\bar \Gamma$ interval,
continues to the $\bar \Gamma - \bar X$ interval,
and then goes to an SS located in the lower local gap at $\bar X$.
The state shows little dispersion as a function of $\bf k_{||}$.
The wavefunction composition of this state has
the $s,d_{x^2+y^2}$ symmetry.
  
An RS was obtained at energies of approximately 3.6 eV 
in the $\bar\Gamma-\bar X$ interval
and seems to have an oscillatory shape.
That is, the state seems to extend throughout
the SBZ, crossing the $\bar X$--point at 3.5 eV,
then crossing the $\bar M$--point at 0.2 eV, and
finally ending at 3.6 eV in the middle of the
$\bar M-\bar \Gamma$ interval.
Although the state seems to be discontinuous in its trajectory,
this could be a
consequence of the numerical accuracy; the state
should be a single band crossing the entire SBZ.
A similar pattern was obtained for the Pt(111) surface.
The wavefunction composition 
of this RS has the $d_{x^2+y^2},d_{xz}$ symmetry.
  
Similar comments are appropriate for the RS that
begins at 2.1 eV in the $\bar\Gamma-\bar X$ interval and seems
to continue through the $\bar X$--point before going
through the $\bar M$--point and mixing with the previously
discussed RS. This state
finally ends at the $\bar \Gamma$--point once again.
Although it is difficult to establish a unique pattern for these
RSs, it could be possible that they represent one band
that crosses the entire SBZ.
The wavefunction compositions found for these RSs are
$d_{xz},d_{z^2},d_{xy}$.
  
At $\bar M$, a lower RS with a parabolic shape as a function of
$\bf k_{||}$ was found. 
This state begins near the local gap located
between 4.0 -- 5.0 eV and ends in the middle of the
$\bar M - \bar \Gamma$ interval.
The wavefunction composition of this RS 
is $s,d_{x^2+y^2}$.
  
Near $E_F$ at the $\bar M$ point, a surface
state with a negative curvature is observed. 
The state goes
into the local gap above $E_F$ with a bandwidth of
approximately 1.1 eV.
The calculated wavefunction composition of this SS 
is $d_{x^2+y^2}$.

\subsection{Comparison with experiment}
  
\begin{table*}
\caption{Calculated energy values and wave function compositions of 
the different surface states found for Pt(001). 
The calculated ETB wave functions are also given for comparison.
For details, see the discussion in the text.}
\begin{tabular}{|c|c|c|c|c|c|}
\hline
\hline
Point & State &
               \multicolumn{2}{|c|} {Energy value (eV)} &
                                    \multicolumn{2}{|c|} {Wave function} \\
\hline
 & & Experiment & Calculated & FLAPW & ETB \\
\hline
\multirow{3}{*}{$\bar\Gamma$} & SS & 5.5\cite{RDrube} & 4.3 & --- & $s,p_z$ \\
 & RS &  & 0.0 & $d_{xz},d_{z^2},d_{xy} $ & $d_{x^2+y^2}$\\
 & RS & --6.5 \cite{APStampfl} & --6.4 & $s,d_{x^2+y^2} $ & $d_{xy},d_{3z^2-r^2}$\\
\hline
\multirow{6}{*}{$\bar X$} & SS & & 4.3 & $s,d_{x^2+y^2}$ & $s,d_{xy}$ \\
 & SS & & 2.1 & --- & $p_x,p_y$ \\
 & RS & & 0.0 & $d_{xz},d_{z^2}$ & --- \\
 & RS & --2.5 \cite{APStampfl} & --2.3 & $d_{xy},d_{xz}$ & --- \\
 & RS & --4.0 \cite{Subaran} & --3.6 & $d_{x^2+y^2},d_{xz}$ & $d$ \\
 & SS & --5.0$\footnote{Value calculated by Benesh {\it et al.}~\cite{Benesh}}$ 
                   & --5.5 & $s,d_{x^2+y^2}$ & $s,d_{xy}$ \\\hline
\multirow{6}{*}{$\bar M$} & RS & & 9.0--10.0 & $d_{xy}$ & --- \\
 & SS & $\sim 0.0$ \cite{APStampfl} & 1.1 & $d_{x^2+y^2}$ & --- \\ 
 & RS & --0.6 \cite{APStampfl} & 0.4 & --- & $d_{xy}$  \\ 
 & RS & --0.9 \cite{APStampfl} & --0.6 & $d_{z^2}$ & ---  \\ 
 & RS &  & --1.7 & --- & $d_{3z^2-r^2}$ \\ 
 & RS & --5.5 \cite{APStampfl} & --5.2 & $s,d_{x^2+y^2}$ & --- \\ 
\hline
\hline
\end{tabular}
\label{tabla2}
\end{table*}
  
It is well known that Pt(100) exhibits both the unreconstructed
$(1\times 1)$ surface and 
the reconstructed $(5\times 1)$ surface
\cite{APStampfl,RDrube,Subaran}. However, 
the ideal surface was studied in this work, 
and the results will be compared with 
experimental data found for the 
$(1\times 1)$ phase.
  
Using angle--resolved photoemission spectroscopy,
Stampfl {\it et al.} \cite{APStampfl} reported the SSs
of Pt(100)$(1\times 1)$
at energies below $E_F$.
These authors present a rich number of SSs along the
$\bar M-\bar\Gamma-\bar X$ interval for the
$(1\times 1)$ phase (see Fig. 2 in Ref.~\cite{APStampfl}).
Although these states are not discussed in detail in
Ref. \cite{APStampfl}, 
it will be shown that the general shape of the reported states is 
reproduced accurately in the present work.
  
As was reported by Stampfl {\it et al.} \cite{APStampfl},
there is an RS near $E_F$ for the $\bar M-\bar \Gamma$ interval
that follows the
border of the $E_F$.
The state shows almost zero dispersion as a function 
of $\bf k_{||}$ (see Fig. 2 in Ref. \cite{APStampfl}).
The DFT calculations found a state around the $\bar M$ point
located mainly in the local gap just above $E_F$, 
and this state 
could be identified with the experimental one.
  
There are two RSs reported at 0.6 and 0.9 eV at the $\bar M$ point.
These states are dispersed throughout nearly the entire 
$\bar M -\bar \Gamma- \bar X$ interval
(see Fig. 2 in Ref. \cite{APStampfl}).
The energy dispersion of these states is worth noting and 
is reproduced properly in the DFT calculations
(see Fig.~\ref{pt100}). 
As mentioned above,
these states show quasi--oscillatory
behavior in this portion of the SBZ.
A similar pattern can also be observed from the calculated
bands shown in Fig. 2(b) in Ref.~\cite{APStampfl}.
  
Stampfl {\it et al.} \cite{APStampfl} reported another 
RS at low energies around 5.5 eV at the $\bar M$ point.
This state is reproduced accurately in the DFT calculation
as discussed above
(see Fig.~\ref{pt100} and Table~\ref{tabla2}).

Near the $\bar X$ point, an RS that reaches the
$\bar X$ point was found around 
2.3 eV.
This state seems to be related to the state around 2.5 eV
reported by Stampfl {\it et al.} \cite{APStampfl}.
  
However,  Subaran {\it et al.} \cite{Subaran}
used angle--resolved photoemission spectroscopy and reported 
a flat band around 4.0 eV at the $\bar X$ point,
which differs from the experimental data
reported in Ref.~\cite{APStampfl}. It was 
speculated that this band represents
emission from the surface layer or that 
it arises from absorbate atoms \cite{Subaran}.
  
As was shown above, our calculations 
found an RS around 3.6 eV that very accurately reproduces the 
dispersion and shape of the state 
reported by Subaran {\it et al.} \cite{Subaran}.
As was mentioned there, this state seems to be 
part of a continuous band that crosses the entire SBZ
(see Fig.~\ref{pt100} and Table~\ref{tabla2}).
  
Stampfl {\it et al.} \cite{APStampfl} reported another RS 
at energies around 6.5 eV at the $\bar \Gamma$ point.
This state exhibits parabolic dispersion as a function of 
$\bf k_{||}$, and as mentioned above, our DFT calculations
properly reproduce this state.
  
Stapfl {\it et al.} \cite{APStampfl} reported an RS 
around --0.3 eV at the $\bar \Gamma$ point, 
and this state is also reproduced in the DFT calculations.
  
It is well known that it is difficult to reproduce
experimental measurements individually. However,
the accuracy of our calculated SSs and RSs
in comparison with those reported by
Stampfl {\it et al.} \cite{APStampfl} 
for the Pt(100) surface is worth noting.

In an early experimental work
Drube {\it et al.} \cite{RDrube} used angle--dependent inverse photoemission, 
To measure the SSs
of Pt(100)$(1 \times 1)$ at energies above $E_F$.
Energy band 
dispersion was found in the $\bar \Gamma - \bar X$ interval.
  
These authors found an SS in the local gap above $E_F$, which is
labeled $S_1$ in Fig. 3 of Ref. \cite{RDrube}.
  
The authors also report a state at 0.6 eV above $E_F$
that seems to be an RS: the state labeled $B_1$ in Fig. 3
of Ref. \cite{RDrube}.
  
They also report a state labeled $D$.
The authors mention that they did not find
an explanation for this state.
  
A state labeled $B_2$, which exhibited significant dispersion, was also
reported. Although this state is found nearly inside the
bulk bands, there is a portion of the state that penetrates into the
local gap near the $\bar X$ point.
  
Discussion of these states and comparison with
our DFT calculations is left for the next section, 
where the ETB results for the Pt(100) surface
will be presented.
  
\subsection{Tight--Binding Calculations}
  
Figure \ref{pt100-tb} shows the calculated pbbs, SSs, 
and RSs for Pt(100) using the FLAPW method
and compares them with those obtained using
the ETB method.
Table~\ref{tabla2} shows the wavefunction compositions of 
the different SSs and RSs.
As in the Pt(111) case,
the ETB calculation properly reproduces the pbbs, 
SSs, and RSs that were found in the DFT 
calculations.
A few discrepancies are observed and will be discussed below.
The observed differences include the fact that 
the ETB calculations do not find the same number of states as
the DFT calculations.
  
Figure \ref{pt100-tb} shows that the local energy gaps found in the ETB
calculations are identical to those calculated using
DFT.
More importantly, the dispersion
of the SSs in the local gaps found by the ETB calculations
is almost the same as those found by the DFT calculations.
  
However, 
the ETB calculations predict
an SS located in the local gap above $E_F$
at the $\bar \Gamma$ point that 
seems to be related to the state that 
was reported by Drube {\it et al.}
\cite{RDrube}. This state was not found in the DFT calculations.
This SS was reported at approximately 5.5 eV,
and the state 
was found at 4.3 eV in the ETB calculation. The state increases in energy to
approximately 6.0 eV and seems to mix with the bulk bands.
The calculated ETB wavefunction composition of this state
exhibits $s,p_z$ symmetry (see Table~\ref{tabla2}).
  
Another SS was found in the 
ETB calculations but not in the 
DFT calculation.
The state exhibits significant dispersion as a function of $\bf k_{||}$,
is located at approximately 2.1 eV 
in the local gap around the $\bar X$ point,
and disperses following the lower edge of the local gap.
The calculated ETB wavefunction composition of this state
has the $p_x,p_y$ symmetry (see Table~\ref{tabla2}).
  
An SS 
following the upper 
edge of the local gap at $\bar X$ was found at approximately 4.3 eV.
It was found that both calculations predict this state,
but no experimental evidence 
for this state was found.
  
At energies below $E_F$, the 
ETB calculation properly reproduces most of the SSs and RSs
found in the DFT calculations,
as shown in Fig.~\ref{pt100}.
In some cases, there are some small numerical 
differences in the calculated
energy values of these states, 
but in general, most of the features found in the 
DFT calculation were also found in 
the ETB calculation.
  
The ETB calculations also reproduce most
of the experimental data reported by Stampfl {\it et al.}
\cite{APStampfl}.
These facts demonstrate the 
predictive power of the ETB method.
  
\section{Platinum(110)}
\label{Pt110}
  
\begin{figure*}[!ht]
\vspace{.75in}
\includegraphics[width=120mm]{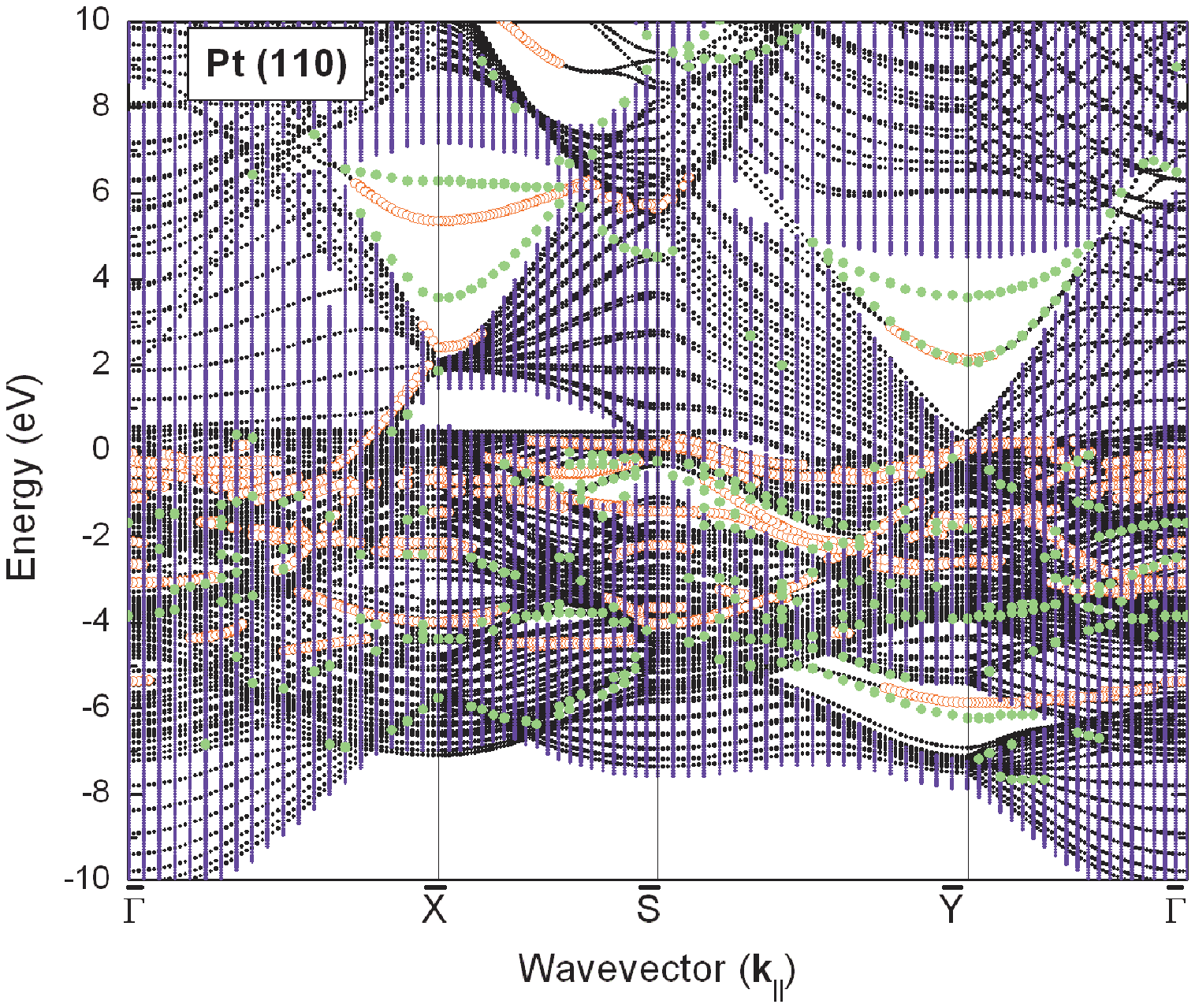}
\vspace{-3cm}
\caption{(Color online) Projected bulk bands of Pt(110).
  The black dots represent the DFT calculated pbbs.
  Red dots represent SSs or RSs if the states
  are located in a local energy gap or in the 
  continuum bulk bands, respectively.
  The blue dots represent the pbbs calculated using the 
  ETB method, and the SSs and RSs are represented 
  by green dots. 
}
\label{pt110}
\label{Pt110-TB}
\end{figure*}

Figure \ref{pt110} shows the calculated pbbs, SSs, and RSs
for Pt(110). 
Table \ref{tabla3} shows the calculated wavefunction
compositions of the different SSs and RSs of 
this surface.
  
As in previous cases, the figure shows the pbbs
as small black dots, and the SSs and RSs
are shown as red dots.
  
The figure shows four local gap above $E_F$, 
and three local gaps are found at energies below $E_F$.
  
Three SSs above $E_F$ were found from the calculations. 
An SS is found in the local gap at the $\bar X$ point around 5.3 eV.
This state exhibits nearly parabolic behavior as a function of
$\bf k_{||}$, and its energy bandwidth is approximately 1.0 eV.
The state mixes with a calculated RS obtained at the $\bar S-$point
at approximately 6.2 eV.
The wavefunction of this SS has $s,p_z$ symmetry.
  
Another SS was located near
the bottom of this local gap.
This state is located at 2.4 eV and 
extends a few $k-$values from the
$\bar X$ point.
The computed wavefunction composition of this state is 
$s,d_{x^2+y^2},d_{xz}$. 
  
Near the $\bar X$ point, there is an RS that should be noted.
This state shows peculiar behavior as a function of $\bf k_{||}$.
The state seems to 
originate in the group of RSs located in the energy range from 0 to 
1.0 eV below $E_F$ and exhibits significant energy dispersion
following the edge of the local gap.
  
An SS was obtained in the local gap at the $\bar Y$ point. 
This state exhibits little dispersion as a function
of $\bf k_{||}$. 
The state is located at approximately 2.1 eV,
and the calculated wavefunction composition of 
this state is $s,d_{x^2+y^2},d_{xz}$. 
  
As for the previous surfaces, 
a number of RSs were found at energies below $E_F$ and are shown in 
Fig. \ref{pt110}.
The main characteristics of these states are as follows:
  
A noticeable SS was found at low energies, approximately 5.9 eV
in the $\bar Y - \bar \Gamma $ interval.
The state begins in the lower local gap at $\bar Y$
and then continues into the continuum of the pbbs in the 
$\bar Y - \bar \Gamma$ interval.
The wavefunction composition of this state is 
$s,d_{x^2+y^2}$.

Similarly, a series of RSs were found near $E_F$ in 
the $\bar Y -\bar \Gamma$ interval
located at 
energies that range from 0.0 to 3.0 eV.
The states then go through the $\bar \Gamma - \bar X$ interval.
  
At energies near $E_F$, around 0.1 eV 
at the $\bar S$ point, an RS was found that
follows the dispersion of the upper pbbs.
This state extends from the middle of
the $\bar X - \bar S$ through the $\bar S - \bar Y$ intervals.
This state is a hybridization of the $s,d_{x^2+y^2},d_{xy}$ 
orbitals.
  
A series of RSs were found at the $\bar X$ point.
There is one RS around 0.7 eV that seems to be part of
the states coming from the $\bar \Gamma - \bar X$ interval and
going to the $\bar S$ point and then to the $\bar Y$ point.
The wavefunction composition of this state is 
$s,d_{yz},d_{z^2}$.
Another RS is located around 2.1 eV.
An RS located at approximately 4.2 eV was also found.
The wavefunction compositions of these states primarily have 
the symmetries of the $d_{xz}$ and $s,d_{xy}$ orbitals,
respectively.
  
A local gap at 0.5 eV is observed at the $\bar S $ point,
and an SS is located there.
The wavefunction composition of this state is primarily
$d_{yz}$.
The already mentioned SS at 1.4 eV was also found at this point 
and shows the wavefunction composition is $d_{xz},d_{z^2}$.
Another RS with a parabolic shape is located around 2.2 eV and
has the wavefunction composition $d_{x^2+y^2}$.
An RS around 4.0 eV was also found. 
The wavefunction composition of this state has the 
$d_{xy}$ symmetry. 
The final RS is located in the $\bar X - \bar S$ interval 
around 4.5 eV. The wavefunction of this state has 
$s,d_{xy}$ symmetry.
  
\subsection{Comparison with experiment}
  
\begin{table*}
\caption{Calculated energy values and wave function compositions of 
  the different surface states of Pt(110).
  For comparison, the calculated ETB wave functions are also given.
  For details, see the discussion in the text.}
\begin{tabular}{|c|c|c|c|c|c|}
\hline
\hline
Point & State &
               \multicolumn{2}{|c|} {Energy value (eV)} &
                                    \multicolumn{2}{|c|} {Wave function} \\
\hline
 & & Experiment & Calculated & FLAPW & ETB \\
\hline
\multirow{3}{*}{$\bar\Gamma$} & SS & & 6.9 & --- &  $s,p_x,p_y$ \\
 & RS &  & $0\to -1.2$ & $d_{yz},d_{z^2}$ & --- \\
 & RS &  & --1.7 & --- & $d_{yz},d_{x^2+y^2}$\\
\hline
\multirow{8}{*}{$\bar X$} & SS &  & 6.2 & --- & $s,p_z $ \\
 & SS & 6.0 \cite{NMemmel}& 5.3 & $s,p_z$ & --- \\
 & SS & & 3.6 & --- & $p_x$ \\
 & SS & & 2.4 & $s,p_{x}$ & --- \\
 & RS & & --0.7 & $s,d_{xy}d_{z^2}$ & --- \\
 & RS & & --2.1 & $d_{xz}$ & --- \\
 & RS & & --4.2 & $s,d_{xy}$ & --- \\
 & RS & & --5.7 & --- & $p_x,d_{xy},d_{yz}$ \\\hline
\multirow{7}{*}{$\bar S$} & RS & & 4.5 & --- & $d$ \\
 & RS & $\sim 0.0$ \cite{AMenzel} & 0.1 & $ s,d_{x^2+y^2},d_{xy} $ & --- \\ 
 & RS &  & --1.4 & $ d_{xz},d_{z^2}$ & --- \\ 
 & RS &  & --2.2 & $d_{x^2+y^2}$ & --- \\ 
 & RS &  & --4.0 & $d_{xy}$ & --- \\ 
 & RS &  & --4.5 & $s,d_{xy}$ & --- \\ 
 & RS &  & --5.2 & --- & $d_{xz}$ \\ 
\hline
\multirow{4}{*}{$\bar Y$} & SS & 5.1 \cite{NMemmel} & 3.6 & --- & $s,p_z$ \\
 & SS & 1.3 \cite{NMemmel} & 2.1 & $s,d_{x^2+y^2},d_{xz}$ & $p_y,d_{yz}$ \\ 
 & RS &  & --3.9 & --- & $d_{yz},d_{x^2+y^2}$ \\ 
 & RS &  & --5.9 & $s,d_{x^2+y^2}$ & $s,p_z,d_{x^2+y^2},d_{3z^2-r^2}$ \\ 
\hline
\hline
\end{tabular}
\label{tabla3}
\end{table*}
  
For energies above $E_F$, 
experimental reports of the electronic band 
structure of this surface can be found \cite{GRangelov, NMemmel}.
To our knowledge, however, no 
studies of the electronic band structure for this surface 
at energies below $E_F$
have been published.
  
It is well established that Pt(110) exhibits a reconstruction called
$(2 \times 1)$ missing row \cite{NMemmel,AMenzel,GRangelov}.
Because an ideal surface calculation was performed here, 
it is not possible to quantitatively compare the results
with the measured values. 
However, the experimental results will be used as a guide to
discuss the calculations.
  
In a recent inverse photoemission (ARUPS) study,
Memmel {\it et al.} \cite{GRangelov,NMemmel}
presented a series of SSs and RSs for the 
$\bar X-\bar\Gamma-\bar Y$ interval.

It is interesting to note that in the local gap
at the $\bar X$ point, Memmel {\it et al.} \cite{NMemmel}
report an SS at approximately 6.0 eV (labeled $S^+_0$ 
in Fig. 3 of Ref. \cite{NMemmel}), 
which is found to be a one--dimensional state.
This result means that the state
is insensitive to the $(1 \times 2)$ missing row reconstruction
\cite{GRangelov,NMemmel}.
  
The one--dimensional character of this state is the
reason that our calculations accurately reproduce this state.
However, the calculated SS shows more dispersion than the
measured state and is predicted at 5.3 eV.

As mentioned above, 
a lower local gap was also calculated at $\bar X$.
The calculations predict that this lower local gap has an energy
width of almost 1.5 eV, whereas the experimental study reports a
gap with an energy width of almost 1.0 eV.
  
At the same time, the calculations predict an
RS that exhibits significant dispersion
along the edge of the lower local gap,
whereas the experimental study presents an RS following the
edge of the upper local gap.
  
In the local gap at the $\bar Y$ point, Memmel {\it et al.} \cite{NMemmel}
report an SS at an energy of 1.3 eV along with
other weak features that should be identified with
umklapp processes from the $\bar \Gamma$ point \cite{NMemmel}.
The DFT calculations found an SS near the lower 
edge of this local gap, at approximately 2.1 eV.
  
At the upper energies, Memmel {\it et al.} \cite{NMemmel} 
report a flat SS at 5.1 eV, labeled $S_0^+$ in Fig. 3 of 
Ref.~\cite{NMemmel}.
However, the DFT 
calculations do not reproduce this state.
  
Just above $E_F$ in the rest of the SBZ, Memmel {\it et al.} \cite{NMemmel}
report several states mixed with the ppbs.
  
The flat state at $E_F$ in the
$\bar X - \bar\Gamma - \bar Y$ interval, 
which should represent an RS, should be noted.
  
There is also a state labeled $C$ that shows a negative
slope centered at $\bar \Gamma$, around 3.0 eV.
  
Finally, there are a series of states around 1.0 eV at $\bar \Gamma$,
shown in Fig. 3 of Ref. \cite{NMemmel}, as well as a state labeled $IS$
around 5.0 eV at $\bar\Gamma$.
  
The DFT calculations do not reproduce these states in detail.
However,  a series of states was found near $E_F$ that
show little dispersion 
as functions of $\bf k_{||}$ 
and covered the $\bar S\bar Y$ interval
(see also the above discussion related to Fig. \ref{pt110}).
  
As discussed above, a number of SSs and RSs were found at energies below $E_F$.
However, because we did not find enough experimental 
data in this energy region,
we only comment on our results for the RS near $E_F$ 
at the $\bar S$ point, and further 
commentaries on the rest of the states will be omitted.
  
The RS at 0.0 eV around the $\bar S$ point was previously discussed by 
Menzel {\it et al.} \cite{AMenzel}. 
These authors mentioned that this state is observed 
in clean Pt(110) surfaces as well as in the 
Br/Pt(110)$-c(2\times 2)$ system. 
In a related work, Minca {\it et al.} \cite{MMinca} also discuss an 
RS at the $\bar X$ point. The authors mention that this state
appears because the bulk energy bands present a flat band along 
the $WLW$ line just below $E_F$. 
This band creates a van Hove singularity at $E_F$.
The bulk band, when projected onto the $\bar S$ 
point of the (110) SBZ, is the origin of the observed resonance state.
The results obtained for the ideal Pt(110) surface show 
that the RS at the $\bar S$ point is a characteristic of this surface 
and is independent of the reconstruction. 
  
Similar observations were noted for the one--dimensional 
SS at the $\bar X$ point above $E_F$, as described 
by Memmel {\it et al.} \cite{NMemmel}.
  
\subsection{Tight--Binding Calculation}
  
Figure \ref{Pt110-TB} shows the pbbs, SSs, and RSs
of the Pt(110) ideal surface found using the ETB method.
In the figure, the blue (black) dots represent the 
pbbs calculated using the ETB (FLAPW) method, while the
green (red) dots represent the SSs and
RSs calculated using the ETB (FLAPW) method. For
details, see the figure caption.
  
Although there are small differences at the edges of
the calculated local gaps above $E_F$,
in general, 
the calculated pbbs at energies below $E_F$
are similar in both methods.
  
At energies above $E_F$, 
a series of SSs were found, and will be 
commented on in detail in the following 
paragraphs.
  
As mentioned above, an SS around 5.3 eV was found at the $\bar X$ 
point in the DFT calculation.
In the ETB calculation, however, an SS
with a quasi--linear shape as a function of $\bf k_{||}$
was found at approximately 6.2 eV.
Although this SS has different energies in the two calculations, 
the wavefunction symmetries bound by the two methods are the same 
(see Table~\ref{tabla3}). 
The state also shows the 
trend reported by Memmel {\it et al.} \cite{NMemmel}.
  
The ETB calculation predicts a second SS around 3.6 eV
at the $\bar X$ point near the lower edge of the local gap.
This state differs in its energy, 
although not in its shape, from the state found at 
2.4 eV in the DFT calculation.
  
The ETB calculation predicts 
an RS at approximately 4.5 eV near the $\bar S$ point. 
The wavefunction composition of this state has 
the full $d$ symmetry.
  
The ETB calculation shows that an RS was found in the 
upper energies around the local gap at the $\bar S$ point, 
around 9.5 eV. However, no experimental evidence 
for this state was found.
The same is true of the SS calculated using the FLAPW method, 
which was located in the upper local gap near the $\bar S$ point 
at approximately 9.0 eV.
  
Two SSs were found in the local gap around the $\bar Y$ point.
The lower state follows the dispersion found in the DFT calculation, 
and the state extends over the entire local gap.
The ETB calculation predicts an upper SS around 3.6 eV 
that was not found in the DFT calculation.
This state could 
be related to the state reported by Memmel {\it et al.} \cite{NMemmel} 
at these energies.
To support this speculation, however, it is necessary to 
assume that this SS is independent of the missing 
row reconstruction.
  
At energies below $E_F$, 
the calculated SSs in the main local gaps
were accurately reproduced in both calculations.
For example, the ETB calculation found an SS around 0.5 eV
in the local gap at the $\bar S$ point
with noticeable dispersion,
in agreement with the state calculated using
the FLAPW method.
  
In the lower local gap at $\bar Y$, the SS found
around 6.2 eV in the ETB
calculation exhibits nearly the same dispersion as the
state found using the FLAPW method.
  
On the other hand, Fig.~\ref{Pt110-TB} shows that
a number of RSs were obtained from the ETB calculations. 
However, most of these RSs do not match any 
states calculated using the FLAPW method.
  
In this case, the two calculations provide us with different
series of RSs, contrary to what was obtained for the Pt(100)
surface (see Fig.\ref{pt100-tb}).
  
A possible explanation of these results could be
the need to include reconstruction effects in
the calculations. 
  
When compared with the experimental data, the ETB
calculation properly predicts the SSs found in the local gaps, 
although some differences in the energies were observed because 
the calculations in this work were for an ideal surface.
Nevertheless, these findings demonstrate the predictive power
of the ETB calculations compared with the more 
computationally demanding methods.
In this sense, the two methods complement each other.

\section{Conclusions}
\label{conclusion}
  
We have calculated the electronic band structure of 
platinum low--index surfaces.
In our calculations, we used both DFT and empirical methods.
From our calculations, we report the
pbbs, SSs, and RSs for ideal Pt(111), 
Pt(100), and Pt(110) surfaces.
Comparisons with experimental data show that our calculations 
properly predict the SSs and RSs for Pt(111) and Pt(100) surfaces.
Because the Pt(110) surface exhibits 
the so--called $(2 \times 1)$  
missing row reconstruction that was not included in our calculations, our 
results compare poorly with the SSs reported for this surface.
However, when the reported SSs are 
independent of the reconstruction, we found that our calculations 
properly reproduce the experimental states.
The results of our calculations for ideal surfaces demonstrate the predictive 
power of the empirical method.

\begin{thebibliography}{20}

\bibitem{JHerreraPt111} H.J. Herrera--Su\'arez, A. Rubio--Ponce,
and D. Olgu\'in, waiting refeere comments.

\bibitem{APStampfl} A.P.J. Stampfl, R. Martin, P. Garner, and
A.M. Bradshaw, Phys. Rev. B {\bf 51}, 10197 (1995).

\bibitem{Subaran} W. Subaran, H. Nakajima, A. Kakizaki, and
T. Ishii,  J. Elec. Spec. and Related
Phenom. {\bf 144--147}, 613 (2005).

\bibitem{RDrube} R. Drube, V. Dose and A. Goldmann,
Surf. Scie. {\bf 197}, 317 (1988).

\bibitem{Benesh} G.A. Benesh, L.S.G. Liyanage, and J.C. Oingel,
 J. Phys. Condens. Matt. {\bf 2}, 9065 (1990).

\bibitem{NMemmel} N. Memmel, G. Rangelov and , WE. Bertel,
Prog. Surf. Scie. {\bf 74}, 239 (2003).

\bibitem{GRangelov} G. Rangelov, V. Dose, Bulg. Chem. Comm. {\bf 26}, 159 
(1993).

\bibitem{AMenzel} A. Menzel, Zh. Zhang, M. Minca, Th. Loerting,
C. Deisl, and E. Bertel, New J. Phys. {\bf 7}, 102 (2005)

\bibitem{MMinca} M. Minca, S. Penner, E. Dona, A. Menzel, E. Bertel,
V. Brouet, and J. Redinger, New J. Phys. {\bf 9}, 386 (2007);
A. Menzel, Z. Zhang, M. Minca, E. Bertel,
J. Redinger, R. Zucca, J. Phys. Chem. Solids {\bf 67}, 254 (2006);


\bibitem{Blaha} P. Blaha, K. Schwarz, G.K.H. Madsen, D. Kvasnicka, J. Luitz,
WIEN2k, An Augmented Plane Wave Plus Local Orbitals Program for Calculating
Crystal Properties, ISBN 3-9501031-1-2, Vienna University of Technology,
Austria, 2001.

%
%


\bibitem{Papa} D.A. Papaconstantopoulos, {\em Handbook of the Band
Strucute of Elemental Solids} (Pleunm, New York, 1986).

\bibitem{Garcia-Moliner}
F. Garc\'ia-Moliner, V. Velasco, {\sl Theory of Single and Multiple
Interfaces} (World Scientific, 1992);
F. Garc\'ia--Moliner and
V. Velasco, Prog. Surf. Sci. {\bf 21}, 93 (1986).

\bibitem{Monkhorst} H.J. Monkhorst and J.D. Pack, Phys.
Rev. B {\bf 13}, 5188 (1976).









\end{thebibliography}
\end{document}